\documentclass[useAMS,usenatbib,amssymb]{mn2e}

\newcommand{\be}{\begin{equation}}
\newcommand{\ee}{\end{equation}}

\title[Elastic properties of polycrystalline dense matter]{Elastic properties of polycrystalline dense matter}
\author[D. Kobyakov and C. J. Pethick]{D. Kobyakov$^{1,2,3}$\thanks{E-mail: kobyakov@theorie.ikp.physik.tu-darmstadt.de
 (DK); pethick@nbi.dk (CJP)} and C. J. Pethick$^{4,5}$\footnotemark[1]\\
 $^{1}$Institut f\"ur Kernphysik, 
Technische Universit\"at Darmstadt, 
D-64289 Darmstadt, Germany\\
$^{2}$ExtreMe Matter Institute EMMI, 
GSI Helmholtzzentrum f\"ur Schwerionenforschung GmbH, 
D-64291 Darmstadt, Germany\\
$^3$Radiophysics Department, Nizhny Novgorod State University, Gagarin Ave. 23, 603950 Nizhny Novgorod, Russia\\
$^4$The Niels Bohr International Academy, The Niels Bohr Institute, University of Copenhagen,\\ Blegdamsvej 17, DK-2100 Copenhagen \O, Denmark\\
$^5$NORDITA, KTH Royal Institute of Technology and Stockholm University, Roslagstullsbacken 23, SE-106 91 Stockholm, Sweden}

\begin{document}

\date{}

\pagerange{\pageref{firstpage}--\pageref{lastpage}} \pubyear{2002}

\maketitle

\label{firstpage}

\begin{abstract}
Elastic properties of the solid regions of neutron star crusts and white dwarfs play an important role in theories of stellar oscillations. Matter in compact stars is presumably polycrystalline and, since the  elastic properties of single crystals of such matter are very anisotropic, it is necessary to relate elastic properties of the polycrystal to those of a single crystal.  We calculate the effective shear modulus of polycrystalline matter with randomly oriented crystallites using a self-consistent theory that has been very successful in applications to terrestrial materials and show that previous calculations overestimate the shear modulus by approximately 28\%.

\end{abstract}


\section{Introduction}
Vibrations of neutron stars and white dwarfs are an important source of information about matter in their interiors.  In recent years the elastic properties of neutron stars have attracted particular attention because vibrations of the crust could provide a mechanism for quasiperiodic oscillations observed in X-ray burst sources \citep{Duncan,Strohmayer}.   The elastic properties of single crystals of dense matter have been calculated in a variety of works (see, e.g., \citet{Haensel, Baiko1, Baiko2}).   However, because of initial variations in the local temperature and composition, as well as the presence of gravitational and magnetic fields, it appears unlikely that the solid part of a star is one giant single crystal, and the standard assumption in recent work is that matter is polycrystalline, with a random distribution of crystal orientations  \citep{Ogata, Baiko1, Baiko2}.  The question we address here is how to calculate the elastic properties of such a polycrystalline medium from those of a single crystal.

%

In white dwarfs and in the crusts of neutron stars at zero temperature matter consists of a lattice of nuclei with a rather uniform background of electrons, which ensure charge neutrality.   In the inner crust of neutron stars, the lattice is also permeated by a sea of neutrons, but these have little influence on the shear properties, which are determined primarily by the Coulomb interaction.

In the presence of a uniform background of electrons, the lowest energy lattice of nuclei with Coulomb interactions has a body-centred-cubic (bcc) structure and its elastic properties are conveniently described in terms of elastic constants $c_{11}$, $c_{12}$, and $c_{44}$ in the standard Voigt notation\footnote{For matter at nonzero pressure, which is the case in compact stars, it is convenient to define elastic constants in terms of variations of the {\it Gibbs} free energy rather than the Helmholtz free energy  since then the formalism at nonzero pressure becomes identical to the standard treatment for zero pressure, as pointed out by \citet{Marcus}.} or alternatively in terms of the quantities $B=(c_{11}+2c_{12})/3$ (the bulk modulus), $c_{44}$, and $c'=(c_{11}-c_{12})/2$, which describe response to three orthogonal distortions of the crystal.  The rigidity of the solid to a shear perpendicular to one of the crystal axes is described by $c_{44}$ and that to a shear perpendicular to the (1,1,0) and equivalent directions by $c'$.    The Coulomb crystal has very anisotropic elastic properties, since the anisotropy ratio \citep{Zener}
\be
{\cal A}=\frac{c_{44}}{c'},
\ee
which is unity for an isotropic medium,
 is approximately 7.5 when screening by electrons is neglected. Therefore, it is important to know how to relate the elastic properties of the polycrystalline medium, which behaves as an isotropic medium at long wavelengths and is therefore described by just two elastic constants, the bulk modulus and an effective shear modulus $\mu_{\rm eff}$.

The earliest estimate of the effective shear modulus of matter in the crust of a neutron star was based on assuming it to be equal to $c_{44}$ \citep{BaymPines}.  More recently, \citet{Ogata} calculated $\mu_{\rm eff}$   for polycrystalline matter with randomly oriented crystallites on the assumption that it is given by the average of the shear stiffness over all possible wavevectors and for polarisation vectors perpendicular to the wave vector, and this prescription has been employed in subsequent work  in which many-body effects are calculated \citep{Baiko1, Baiko2}.  For a cubic crystal, one finds
\be
\mu_{{\rm eff}, V}=\frac25 {c'} +\frac35 c_{44},
\label{voigt}
\ee
where the subscript $V$ is chosen since this way of averaging was proposed earlier by  \cite{Voigt}, a fact pointed out by \citet{Johnson-McDaniel} in the context of the shear properties of mixed phases of quark matter and hadronic matter at supernuclear densities.

\section{Effective shear modulus}

The elastic properties of polycrystalline terrestrial materials is a subject of considerable practical interest and there exists an extensive literature on the subject.  We now give a brief overview.  \citet{Reuss} proposed that, instead of averaging the shear rigidity, one take its harmonic mean.  This corresponds to taking the shear compliance of the polycrystal, the inverse of the shear rigidity, to be the average over wavevectors and polarisations of the shear compliance of the single crystal. For a cubic crystal this leads to
\be
\mu_{{\rm eff}, R}=\left(\frac25\frac1{c'} +\frac35\frac1{ c_{44}}\right)^{-1}.
\ee
For  the case of a Coulomb crystal, the Voigt and Reuss prescriptions differ by a factor of about 2.5, so it is important to narrow down the uncertainty in calculating $\mu_{\rm eff}$.  \citet{Hill} showed that, with the assumption that the crystallites have random orientations, the Voigt average gives an upper bound to $\mu_{\rm eff}$ and the Reuss average a lower bound.  He proposed that for a more realistic estimate one should use either the arithmetic mean of them, which is given by
 \be
\mu_{{\rm eff}, H}=\frac12(\mu_{{\rm eff}, V}+\mu_{{\rm eff}, R}),
\ee
which has been commonly used in subsequent work, or the geometric mean $(\mu_{{\rm eff}, V}\mu_{{\rm eff}, R})^{1/2}$.
Following that, \citet{Hershey}, \citet{Kroener} and \citet{Eshelby} developed a model in which one calculates the strain field around an inclusion (representing a single crystallite) in a homogeneous medium and then, by averaging over the possible orientations of the crystallite, determines the effective shear modulus self-consistently (for a pedagogical account of the theory, see \citet{deWit}).  In the self-consistent theory one finds
\be
8 \mu_{\rm eff}^3 +(9B+4\,{c'})\mu_{\rm eff}^2
-3(B+4\,{c'})\,c_{44}\,\mu_{\rm eff}-6B\,{c'}\,c_{44}=0.
\label{muSC}
\ee
The predictions of the self-consistent theory are very robust, since very similar values of the effective shear modulus are obtained from a rather different approach in which one treats the medium as an elastic medium with smooth spatial variations of the elastic properties \citep{Kroener2}.

\subsection{Comparison with experiment}

The self-consistent theory gives a rather good account of experimental measurements of elastic properties of terrestrial materials.  The anisotropy of matter at high density is extreme, and there are rather few materials with such high values.  We shall restrict ourselves to cubic materials.

 One system that has been studied extensively is the $\delta$-phase of plutonium,\footnote{This phase has a face-centred-cubic structure.  However, since the theories of average elastic constants depend only on the cubic symmetry and not on the detailed arrangement of atoms within a cubic cell, this is a relevant system for comparison with theory.  This phase may be stabilised at room temperature by addition of a small amount of gallium, and the results we quote are for this dilute alloy.} for which ${\cal A}\approx 7.03$ \citep{Ledbetter}.  The experimentally determined shear elastic constant, 16.3 GPa \citep{Migliori},  agrees well with the result of the self-consistent theory for the single-crystal elastic constants measured by \citet{Ledbetter}, which is 16.2 GPa. Two other materials  with significant elastic anisotropy that have been studied in detail are copper, for which ${\cal A}\approx 3.14$  \citep{LedbetterCu}, and stainless steel, with  ${\cal A}\approx   3.78$ \citep{LedbetterSS}: in both these cases, the self-consistent theory gives results that agree with experiment to within a few percent.

The alkali metals also have high elastic anisotropy and bcc structure.  However, these substances are difficult to work with and we have been unable to find recent measurements of the elastic constants of polycrystalline alkali metals with well characterised samples.  Lithium has  ${\cal A}\approx  8.52 $, and the self-consistent theory predicts $\mu_{\rm eff}=4.2$ GPa \mbox{\citep{deWit}}, while the experimental value is 4.0 GPa.   The behaviour of sodium (${\cal A}\approx 6.98$) appears to be anomalous:  the measured shear elastic constant quoted in compilations of data is 3.3 GPa,  while the self-consistent theory predicts 2.1 GPa.  What is even more surprising is the fact that the experimental value is greater than an upper bound in the theories, the Voigt average Eq.\ (\ref{voigt}), which is  2.8 GPa.
Without more information about the structure of polycrystalline sodium, the reasons for this discrepancy are unclear.

\section{Astrophysical dense matter}

We now apply the results of the previous section to dense stellar matter.  For illustrative purposes we shall restrict ourselves to the case of zero temperature and will neglect many-body effects.
For a Coulomb lattice with a uniform background of negative charge, the elastic constants are given by
\citep{Fuchs,Baiko1}\footnote{The elastic constants used in the present article are related to the quantities in \citet{Baiko1} by $c'=(S_{1111}-S_{1122}-P)/2$ and $c_{44}=S_{1212}$, where $P$ is the pressure.}
\be
c'=  0.09947\frac{n_N Z^2 e^2}{2a}
\ee
and
\be
c_{44}=  0.7424\frac{n_N Z^2 e^2}{2a},
\ee
where $a$ is the length of the side of a cubic cell containing two nuclei, $n_N$ is the number density of nuclei, $Z$ is the proton number of the nucleus, and $e$ is the elementary charge.

The Voigt and Reuss predictions for the effective shear modulus of a polycrystal are therefore
\be
\mu_{{\rm eff}, V}=0.4852\frac{n_N Z^2 e^2}{2a}
\ee
and
\be
\mu_{{\rm eff}, R}=0.2071\frac{n_N Z^2 e^2}{2a},
\ee
while Hill's suggestion gives
 \be
\mu_{{\rm eff}, H}=0.3462\frac{n_N Z^2 e^2}{2a}.
\ee

We turn now to the self-consistent theory.  In contrast to the Voigt, Reuss, and Hill expressions, the expression (\ref{muSC}) for the shear modulus involves the bulk  modulus in addition to the two shear elastic constants of the single crystal.  In solid matter in compact stars, the bulk modulus is dominated by that of the elections, $n_e\partial P_e/\partial n_e$, where $n_e=Zn_N$ is the electron density and $P_e$ the electron pressure.  This is of order $n_e^{4/3}\hbar c$ for relativistic electrons and is greater than the shear elastic moduli by a factor $\sim 1/(\alpha Z^{2/3})\ga 10$,  where $\alpha=e^2/\hbar c \approx 1/137$ is the fine structure constant.  Thus it is a good approximation in Eq.\ (\ref{muSC}) to take only the terms linear in $B$.   This gives
\be
3\mu_{\rm eff}^2-\,c_{44}\,\mu_{\rm eff}-2\,{c'}\,c_{44}=0,
\ee
 for which the only physically meaningful root (positive $\mu_{\rm eff}$) is
 \be
 \mu_{\rm eff}=\frac{c_{44}}{6}\left(1+\sqrt{1+24 \frac{c'}{c_{44}}}\right).
 \ee
For the Coulomb lattice,
\be
\mu_{\rm eff}=0.3778\frac{n_N Z^2 e^2}{2a}.
\ee
The Voigt average previously used in the astrophysical literature thus overestimates the shear modulus by approximately 28\%. This effect is much greater than a number of many-body effects, such as electron screening.

\section{Observational considerations and concluding remarks}
Our results show that, in neutron stars, the method previously used to calculate the effective shear modulus overestimates  the frequencies of torsional modes of the crust  by $\sim 15\%$  when crystallites have random orientations.

In white dwarfs,  shear waves in the solid will be similarly affected. However, because of the acoustic impedance mismatch between the solid and liquid regions of the stars, such modes couple weakly to the surface layers of the star, and are consequently not easily detectable by observations of the stellar surface  \citep{Montgomery}.

The self-consistent theory can be applied to quantities other than the shear modulus.  For the bulk modulus \citep{Hershey} and thermal expansion coefficient \citep{Tome}, it predicts for cubic materials that they are the same as those of the single crystal.

The polycrystalline nature of the matter will affect the phonon spectrum, since for wavelengths longer than the typical dimension of a crystallite, the  spectrum is that of an isotropic solid while for shorter wavelengths it is well approximated by the result for a single crystal, provided the size of a crystallite is large compared with the lattice spacing. For example, the phonon heat capacity will be proportional to $T^3$ for temperatures low compared with the Debye temperature but the coefficient will be different for  $T\ga \hbar v_\perp/(L k_B)$ and  $T\la \hbar v_\perp/(L k_B)$, where $L$ is the typical dimension of a crystallite, $k_B$ is the Boltzmann constant, and $v_\perp$ is a typical velocity for transverse sound. 

 Throughout our discussion we have assumed that crystallites are oriented randomly: further work is required to determine whether or not this is a good assumption.    This requires an investigation of the conditions prior to crystallisation, including temperature inhomogeneities, and nuclear composition, as well as the possible role of gravitational and magnetic fields.  Such studies are also important for determining the typical dimension of a crystallite.

A further assumption we have made is that all nuclei have the same atomic number $Z$.  If the dispersion in $Z$ is large, the assumption that matter may be regarded as a collection of perfect crystallites could be a poor starting point.   

In this Letter we have confined our attention to densities at which nuclei are spherical to a good approximation.  In the pasta phases, where nuclei are highly aspherical the calculation of elastic properties is more complicated since it requires a generalisation of the self-consistent theory to liquid-crystal-like systems.

\section*{Acknowledgments}

We thank Albert Migliori, G\"oran Grimvall, and Don Winget for helpful correspondence.  We have had useful conversations with Don Lamb, Paul Steinhardt and  Sidney Nagel. Kader Ahmad was helpful in obtaining copies of older articles.     This work was completed while we enjoyed the  hospitality of the International Space Science Institute, Bern, and ECT*, Trento. This work was supported by the ERC Grant
No.~307986 STRONGINT, by the Helmholtz Alliance Program of the
Helmholtz Association, contract HA216/EMMI ``Extremes of Density and
Temperature: Cosmic Matter in the Laboratory'', and by NewCompStar, COST Action MP1304.

\end{document}